# Metrics for Assessing The Design of Software Interfaces

Hani Abdeen, Osama Shata

hani.abdeen@qu.edu.qa     --     sosama@qu.edu.qa
Computer Science Department, College of Engineering – Qatar University , Doha, Qatar

**ABSTRACT**: **Recent studies have largely investigated the detection of class design anomalies. They proposed a large set of metrics that help in detecting those anomalies and in predicting the quality of class design. While those studies and the proposed metrics are valuable, they do not address the particularities of software interfaces. Interfaces define the contracts that spell out how software modules and logic units interact with each other. This paper proposes a list of design defects related to interfaces: shared similarity between interfaces, interface clones and redundancy in interface hierarchy. We identify and describe those design defects through real examples, taken from well-known Java applications. Then we define three metrics that help in automatically estimating the interface design quality, regarding the proposed design anomalies, and identify refactoring candidates. We investigate our metrics and show their usefulness through an empirical study conducted on three large Java applications.**

## I. INTRODUCTION

Interfaces are the main tool for information hiding in software systems that have them: they represent service contracts between users and providers of behavior. Because of this contract role, interfaces are different from classes, even abstract: they should be more stable across the evolution of a software system, to help reduce the effort to understand and maintain a software system [23]. Designing an interface is a sensitive task with a large influence on the rest of the system. Similarly, during the evolution of a software system, the design of interfaces must be assessed precisely in order to control the impact of any required change [23], [7].

However, as software evolves over time with the modification, addition and removal of new classes and services, the software gradually drifts and looses quality [10]. To help maintainers improve the software quality, there has been recently an important progress in the area of software automatic refactoring and optimization of code quality [19]. Most of the existing approaches in that field are mainly based on code metrics, such as metrics defined by Chidamber and Kemerer [6], and predefined bad smells in source code, by Fowler and Beck [12]. In spite of this progress, unfortunately, none of those approaches takes into account the particularities of interfaces –since interfaces do not contain any logic, such as method implementations, invocations, or attributes.

In the literature, few recent works attempt to address the particularities of interfaces. There are some well-known interface design principles, like Dependency In-version "Program to an interface, not an implementation" or Interface Segregation "Do not design fat interfaces" [18], [13], [26]. However, most existing design patterns, code smells, and metrics revolve around classes without focusing on the specifics of interfaces [23]. To the best of our knowledge, there exist few publications focused on the interface design quality. Besides Martin's design principles ISP and DIP [18], Boxall and Araban's count-ing metrics [4], and the Service Interface Usage Cohesion (SIUC) metric proposed by Perepletchikov [22], there are no tools that help estimate the interface design quality and detect design anomalies in software interfaces.

In this paper, we identify and describe, through examples taken from well known software systems, two design defects specific to software Interfaces:

- Similarity between interfaces: the case of interfaces that represent redundant declarations of service con-tracts between users and providers of behavior.
- Redundancies in interface hierarchies: the case of interfaces that are specified as super-types multiple-times, directly and/or indirectly.

Then, we define a list of metrics that assist in evaluating the design quality of software interfaces, with regard to those design defects. We investigate our metrics through a case study of three open-source Java applications: *JBoss, Vuze* and *Hibernate*. The study shows that the proposed design defects are present, to different degrees, at interfaces. It also shows that the proposed metrics are useful for estimating the quality of software interface design and identifying refactoring candidates.

The remainder of the paper is organized as follows. In Section II we introduce the vocabularies we use in this paper. We identify two interface-specific design defects in Sections III and IV, and define metrics that help in assessing the design of interfaces and in locating those anomalies in code. We then evaluate our metrics by applying them to three Java applications in Section V. Before concluding, Section VI lists existing works related to assessing the design of interfaces.

## II. VOCABULARY

Interfaces are used to encode similarities which the classes of various types share, but do not necessarily constitute a class relationship. Interfaces are usually used to define contracts that spells out the interactions between software modules. In such a context, Interfaces are also meant to define the Application Programming Interfaces (APIs). Hence, Interfaces are supposed to avoid contract violation and to reduce the effort to understand and maintain a software system [23].

**Interface.** We consider an interface as a set of signatures (*i.e.,* method declarations). In this paper we consider only interfaces that declare at least one method, explicitly *"public"*. Thus we do not take into account Marker interfaces or interfaces that are used to declare only constants. We define the size of an interface $i$ as the number of public methods that $i$ declares. We use $i_{size}$ to denote the size of $i$.

**Signature (Method Declaration).** We consider an interface *"signature"* as a *"service"*, to be provided by the interface implementing class(es). We use $sig$ to denote an interface signature, $i_S$ to denote the set of signatures declared in $i$. We say that two signatures are identical if they have the same *return-type*, the same *name* and the same *list-of-parameter-types*.

**Sub-Hierarchy.** We define the sub-hierarchy of an interface $i$ as the collection of all hierarchies of the direct sub-classes and interfaces of $i$. We use $dSub(i)$ to denote the set of classes and interfaces that *directly* specify the interface $i$ as a super-type. We also use $withSubH(x)$ to denote the collection of classes and interfaces that implement or extend, directly or indirectly, $x$, in addition to $x$ itself; where $x$ is either a class or an interface: $withSubH(x) = \sqcup_y withSubH(y) : \forall y \in dSub(x)$. Thus, we define the sub-hierarchy of an interface $i$ as follows: $subH(i) = \sqcup_x withSubH(x) : \forall x \in dSub(i)$. Due to the Interface multiple inheritance mechanism, $subH(i)$ may contain repeated elements.

## III. INTERFACE SIMILARITY AND API DUPLICATION

Interfaces are usually used to define reference types that encode similarities among classes. Software inter-faces are also meant to represent service contracts between users and providers of behavior [26]. Thus, they are also used to define the system APIs [25]. However, software systems evolve over time to add new features and services, to adapt to the environment changes, etc. [10]. As a consequence, code decays and the organization of software interfaces gradually drifts [16]. Code clones are one of the best known bad smells in source code [12], [2]. Although interfaces do not provide implementations, code clones may still occur by duplicating method declarations in several interfaces. Such interfaces that share redundant signature declarations are thus similar from the point of view of public services/APIs they specify. Hence, they indicate a bad organization of the APIs.

### A. Example on Interface Similarity

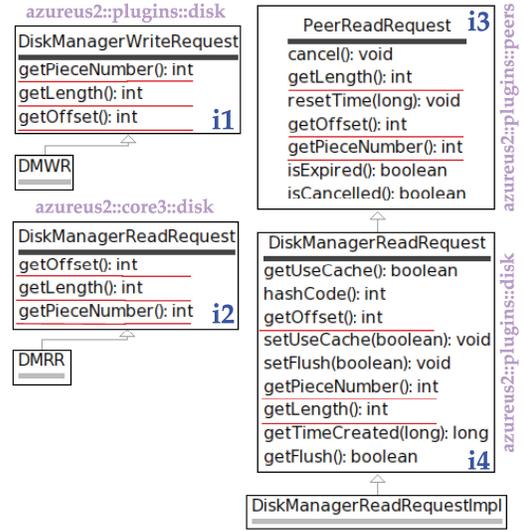

Fig. 1 Example of Interface Similarity. Duplication of read & write request services in four interfaces in *Vuze*.

Fig. 1 shows an example of interface similarities, taken from *Vuze*. It shows a group of three signatures related to read&write request services that are duplicated in four different interfaces. Moreover, the figure shows that two interfaces ($i_1$ and $i_2$) declare the exact same set of those duplicated services ($DS$): 1) getPieceNumber():int; 2) getLength():int; 3) getOffset():int. The interfaces declare those services are: $i_1$ - *DiskManagerWriteRequest*, in *plugins::disk* package; $i_2$ - *DiskManagerReadRequest* (in *core3::disk* package); $i_3$ - *PeerReadRequest* (in *plugins::peers* package); $i_4$ - *DiskManagerReadRequest* (in *plugins::disk* package).

A conclusion is that to locate the read & write request methods in *Vuze*, a developer needs to inspect four interfaces and their different sub-hierarchies (*i.e.,* implementation classes). Which is not the case if those methods were declared only once in one interface. As a refectoring example for reducing the declaration redundancy of those duplicate methods in *Vuze*, we propose to replace interface $i_1$ by interface $i_2$. Thus we should move all the dependencies pointing to interface $i_1$ to interface $i_2$: *e.g.,* make the class *DMWR* implement $i_2$ instead of $i_1$. As a consequence, we can safely remove interface $i_1$ from *Vuze* application. It is worth to note that the implementation classes of $i_1$ and $i_2$ provide identical implementations for those read&write request methods. Then we propose to make interface $i_3$ a sub-interface of interface $i_2$; and remove the set of those duplicate methods from both $i_3$ and $i_4$. The result of our proposed refactoring is that: the read&write methods are declared only in interface $i_2$; the number of interfaces is reduced since interface $i_1$ is removed; and the size of interfaces $i_3$ and $i_4$ is reduced.

## B. Interface Similarity Metrics

To assist software maintainers in automatically quantifying the similarity among interfaces and detect interface clones, we define two new metrics:

- **Index of Interface Similarity** (IIS): this metric aims at quantifying the similarity between interfaces, with regard to the APIs/contracts that those interfaces specify.
- **Index of Interface Clone** (IIC): the IIC metric is a slightly modified version of IIS. It aims at determining the extent to which a given interface is duplicated (cloned) by other interfaces.

By definition, two interfaces $i_1$ and $i_2$ are completely similar, from contracted services perspective, if they declare exactly the same set of methods. Thus, we say that there exists a similarity between two interfaces $i_1$ and $i_2$ if both declare identical signatures: $i_{1_S} \cap i_{2_S} \neq \phi$. We then say that the similarity degree between $i_1$ and $i_2$ is proportional to the size of the shared set of signatures between $i_1$ and $i_2$. We define the Interface Similarity ($IS$) between an interface $i_1$ and another one $i_2$ as follows:

$$IS(i_1, i_2) = \begin{cases} \frac{|i_{1_S} \cap i_{2_S}|}{|i_{1_S} \cup i_{2_S}|} & : i_{1_S} \neq \phi \wedge i_{2_S} \neq \phi \\ 0 & : else \end{cases} \quad (3.21)$$

$IS(i_1, i_2)$ takes its values in [0..1], where 1 is the worst value. The largest value $IS(i_1, i_2)$ has, the largest the similarity between both interfaces. $IS(i_1, i_2)$ quantifies the similarity between two interfaces, with regard to the method declarations in both of them. To determine the set of interfaces that are candidate of refactoring because of their similarities, with regard to a given interface $i$, we use $simi(i)$ to denote the set of all interfaces that have any similarity degree with $i$:

$$simi(i) = \{x \mid x \in \mathcal{I} \wedge IS(i, x) > 0\} \quad (3.22)$$

However, since all the interfaces in $simi(i)$ have certain similarity with $i$, as a consequence, each of them have certain similarity with others. To determine the extent to which a given interface $i$ is similar to other interfaces in the concerned application, and to determine the organizational quality of services into interfaces, we define the **Index of Interface Similarity (IIS)**, for an interface $i$ as well for all interfaces $\mathcal{I}$, as follows:

$$IIS(i) = \begin{cases} \max_x IS(i, x) & : \forall x \in semi(i) \\ 0 & : semi(i) = \phi \end{cases} \quad (3.23)$$

$$IIS(\mathcal{I}) = \frac{\sum IIS(i)}{|\mathcal{I}|} \qquad : \forall i \in \mathcal{I} \quad (3.24)$$

$IIS(i)$ returns the maximal (worst) similarity that $i$ has with any interface in the studied application. $IIS(i)$ takes its values in [0..1], where 0 means that the services/contracts that are described by $i$ are well organized since none of them is re-declared in another interface –*i.e.,* since there is no interface similar to $i$.

The largest value IIS($i$) has, the largest the similarity between $i$ and certain interface(s). For example, supposing that the value of IIS($i$) is 0.75. This means that there is at least one different interface, $x$, where 75% of methods declared in both interfaces $i$ and $x$ are identical. As a consequence, $i$, and its similar interfaces, are candidate for refactoring to reorganize the services/APIs described in $i$. IIS($\mathcal{I}$) aims at determining if the software interfaces organize well the APIs and shared services between classes or not. It quantifies the quality of services/APIs organization into all software interfaces.

A specific scenario of interface similarity is that when some interfaces are completely duplicated/cloned into some of their similar interfaces. In such a case, the software developer should improve the re-usability of interfaces and remove those hard copies (*i.e.,* clones). He/she should use the inheritance mechanism between interfaces instead of hard copying some interfaces into other ones. To measure the extent to which a given interface $i_1$ is cloned within another interface $i_2$, we slightly modify the IS($i_1, i_2$) metric ((3.21)) to define the Interface Clone ($IC$) metric as follows:

$$IC(i_1, i_2) = \begin{cases} \frac{|i_{1_S} \cap i_{2_S}|}{|i_{1_S}|} & : i_{1_S} \neq \phi \wedge i_{2_S} \neq \phi \\ 0 & : else \end{cases} \quad (3.25)$$

$IC(i_1, i_2)$ takes its values in [0..1], where 1 is the worst value. $IC(i_1, i_2)$ quantifies the extent to which $i_1$ is duplicated into $i_2$. The largest value $IC(i_1, i_2)$ has, the largest the subset of $i_1$' method declarations are duplicated in $i_2$. Similarly to IIS metrics ((3.23) and (3.24)), to determine the extent to which a given interface $i$ is *cloned* into other interfaces, and measure the reusability of interfaces, we define the **Index of Interface Clone (IIC)**, for an interface $i$ as well for all interfaces $\mathcal{I}$, as follows:

$$IIC(i) = \begin{cases} \max_x IC(i, x) & : \forall x \in semi(i) \\ 0 & : semi(i) = \phi \end{cases} \quad (3.26)$$

$$IIC(\mathcal{I}) = \frac{\sum IIC(i)}{|\mathcal{I}|} \qquad : \forall i \in \mathcal{I} \quad (3.27)$$

IIC($i$) takes its values in [0..1], where 1 is the worst value and it means that $i$ is completely cloned within certain interfaces. The largest value IIC($i$) has, the largest the clone of $i$. IIC($\mathcal{I}$) aims at quantifying interface clones within the software application under analysis.

## IV. REDUNDANCY IN INTERFACE HIERARCHY

Multiple class inheritance mechanism has several disadvantages related to conflicting features, accessing overridden features, and factoring out generic wrappers [9]. The most known disadvantage of multiple class inheritance is the *diamond problem* which arises when a class inherits from the same base class via multiple paths [5]. One of the advantages of Java interfaces is that they solve the diamond problem [25]. On the other hand, java multiple inheritance mechanism via interfaces has its disadvantages. A particularly structural disadvantage of multiple interface inheritance is that classes may implement a given interface several times, directly and not-directly. This is nothing else than additional static dependencies without any add value; where programmer might struggle to understand the Interface hierarchy. Such useless redundancy in the Interface sub-hierarchy add difficulty to understand the interface hierarchy, and useless dependencies among subsystems [1, 11]. We mainly relate such a design anomaly to the Don't Repeat Yourself (DRY) principle [15]: *"every piece of knowledge must have a single, unambiguous, authoritative representation within a system"*.

### A. Example on Redundancy in Interface Hierarchy

Fig.2 shows an example about redundancy in interface hierarchy. It shows the sub-hierarchy of interface *WorkUnitMergeDispatcher* interface (*i*), in package *envers::synchronization::work*, from *Hibernate*. This interface is extended by interface *AuditWorkUnit* (*ii*). This latter declares 8 services/signatures, representing the shared services of every *WorkUnit* class in *Hibernate*: the classes annotated by *b, c, ..., g*. In fact, all those classes are sub-classes of class *AbstractAuditWorkUnit* (*a*). The class a implements the interface *AuditWorkUnit* (*ii*) and provides a partial implementation of *ii* services. As the classes *b, c, ..., g* are all sub-classes of *a*, they are then, by convention, sub-types of the interface *ii*. Thus, they have to complete the missed implementation of *ii* in *a* -even if they are not declared explicitly as implementation of *ii*. Fig.2 shows that all those classes declare explicitly the implementation of interface *ii*. As a consequence, they add 6 static dependencies pointing to *ii*, which are simply repetitions to what is sated by sub-classing a. As a consequence example, each time a programmer need to modify the interface *AuditWorkUnit* (*ii*) ( *e.g.,* re-naming it or replacing it with another interface), he/she must insure that the refactoring is applied on all those 7 classes. This is 7 times more than applying the modifications on the class a only.

As a summary, we state that the repetitions of interface sub-typing cause additional, needless, static coupling among system classes.

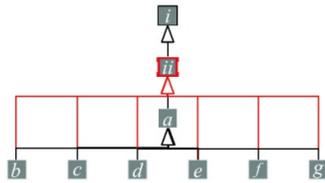

Fig. 2 The sub-hierarchy of *WorkUnitMergeDispatcher* interface ( *i* ), in *envers::synchronization::work* namespace –from *Hibernate*. Examples about redundancy in Interface sub-hierarchy. *Annotations: i, ii, ...* denote Interfaces. *a, b, ..., g* denote classes.

### B. Index of Redundancy in Interface sub-Hierarchy

To determine the degree of redundancy within an interface sub-hierarchy, we simply compute the number of useless repeats of inheritance/implementation relationships within the interface sub-hierarchy. We define the sub-hierarchy of an interface *i* as the collection of all possible sub-hierarchy paths having *i* as a root. Supposing that *i* has *X* direct sub- classes and/or interfaces ( $X = dSub(i)$ ). Then the sub-hierarchy of *i* is the union of *X* with all the sub-hierarchy of elements in *X*. We define the Index of Redundancy in Interface Hierarchy (IRIH) as proportional to the number of repetitions within the interface sub-hierarchy. Let $occurrences(x,i)$ denotes the occurrences of the node *x* in the *i* sub-hierarchy. Let $rep(x,i)$ denotes the duplication occurrences of *x* within $subH(i)$ :

$$rep(x,i) = occurrences(x,i) - 1$$

Let $rep(i)$ denotes the sum of all repetitions of elements in $subH(i)$ :

$$rep(i) = \sum_x rep(x,i) \, , \forall x \in subH(i)$$

Then, the IRIH for an interface *i* is defined as follows:

$$IRIH \ (i) = \frac{rep \ (i)}{\mid subH \ (i) \mid} \quad subH \ (i) \neq \phi \ (4.21)$$

## V. OUR METRICS IN PRACTICE

To investigate our metrics, we applied them to 3 well known and large Java open-source applications: *JBoss*, *Vuze* and *Hibernate*. We chose those applications since they are widely used by the open-source community and contain a large number of interfaces. They also differ in terms of: utility and services, number of interfaces, interface size and class/interface hierarchy (Table I).

We obtained those applications from sourceforge.net. We used the platform Moose for data and software analysis [8] to parse the application source-code and compute the values of our metrics: IIS, IIH and IRIH. Please note that the information we show is this paper about the studied applications are obtained after excluding the following: Java library interfaces and classes; 'constraint & marker' interfaces (*i.e.,* $i_{size}$ = 0); and test-case classes –*i.e.,* classes inheriting from '*JUnit TestCase*' class, or packaged into 'test(s)' packages. The following section provides a short description of the case-study applications.

### A. Case Studies Overview

TABLE I
INFORMATION ABOUT CASE-STUDY APPLICATIONS

| System | $\mid C \mid$ | $\mid I \mid$ | $\mid C_I \mid$% | $i_{size}$ | | |
|---|---|---|---|---|---|---|
| | | | | min | max | sum |
| *JBoss* AS | 5990 | 1414 | 15% | 1 | 75 | 6835 |
| *Vuze* | 6564 | 1020 | 40% | 1 | 195 | 6922 |
| *Hibernate* | 6195 | 541 | 18% | 1 | 354 | 3502 |

$\mid C \mid$ denotes the number of classes (not interfaces); $\mid I \mid$ denotes the number of interfaces; $\mid C_I \mid$% denotes the percent of classes that implement directly interfaces; *sum* $i_{size}$ denotes the number of all declared public methods ( *i.e.,* APIs/services) in interfaces.

### B. IIS and IIC Metrics

In this section we discuss the values of our metrics IIS and IIC in the context of the studied applications: *JBoss, Vuze & Hibernate.*

1) Overview on Interface Similarities and Clones in Studied Applications: Fig. 3 shows an overview on inter- face similarities and clones in the studied applications *JBoss, Vuze* and *Hibernate.* It surprisingly shows that the average similarity between all the software interfaces, small and large ones, in all case-studies is notifiable, where interface clones seem to be considerably present, particularly in *JBoss* and *Vuze* applications. This may indicate that the organization of APIs/services into inter- faces, as well the reusablity of software interfaces, should be investigated to improve interface design quality. For further information on interface similarity and clones, Fig. 4 shows the spread of IIS($i$) and IIC($i$) values over the interfaces of *JBoss, Vuze* and *Hibernate.* It shows the metric values regarding interfaces of differ- ent sizes: the vertical axes represents the metric values; the horizontal axes represents interface sizes (interfaces are sorted on the horizontal axes by their sizes, in ascandant way); red "dots" represent the values of IIC; blue "crosses" represent IIS values.

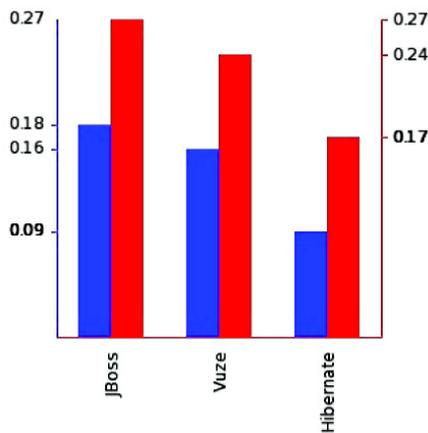

Fig. 3 Overview on Interface Similarity and Clones. IIS( $i$ ) (blue) and IIC( $i$ ) (red) values for the case-study applications.

2) *Many interfaces share sets of method declarations:* Fig. 4 shows that the values of IIS($i$) (blue crosses) and/or IIC($i$) (red dots) spread over a relatively large set of interfaces, of different sizes, and over the interval [0..1], whatever in *JBoss, Vuze* and/or *Hibernate.* This shows that many interfaces share sets of method declarations with other interfaces. Fortunately, a relatively large part of the software interfaces do not have considerable values of IIS and/or IIC –*i.e.,* are not noticeably similar to (and/or cloned in) other interfaces. However, the figure shows that there are many interfaces that have an important similarity with other interfaces, and/or a large partition of their method declarations are cloned –see in Fig. 4 the IIS and IIC values that are surrounded by a dashed rectangle and annotated by 'suspect interfaces'. Those interfaces outline a bad organization of APIs/services within the interfaces.

3) *Very similar interfaces (suspect interfaces):* let us consider the value 0.5 as a critical value of IIS and IIC. This value defines the base borderline for interfaces that more than the half (50%) of their methods are shared/cloned with/in other interfaces. Fig. 4 shows that unfortunately, in all case studies, there are many interfaces that are very similar to other interfaces in the concerned system, and/or a large part of their methods are duplicated in larger interfaces. Such interfaces are suspected in considerable redundancy of APIs/services and bad reusability. Regarding interfaces that have critical values of IIS (*i.e.,* IIS($i$) > 0.5), this means that those interfaces are very similar between each other: *i.e.,* more than 50% of the APIs/services of each of them are shared with another interface.

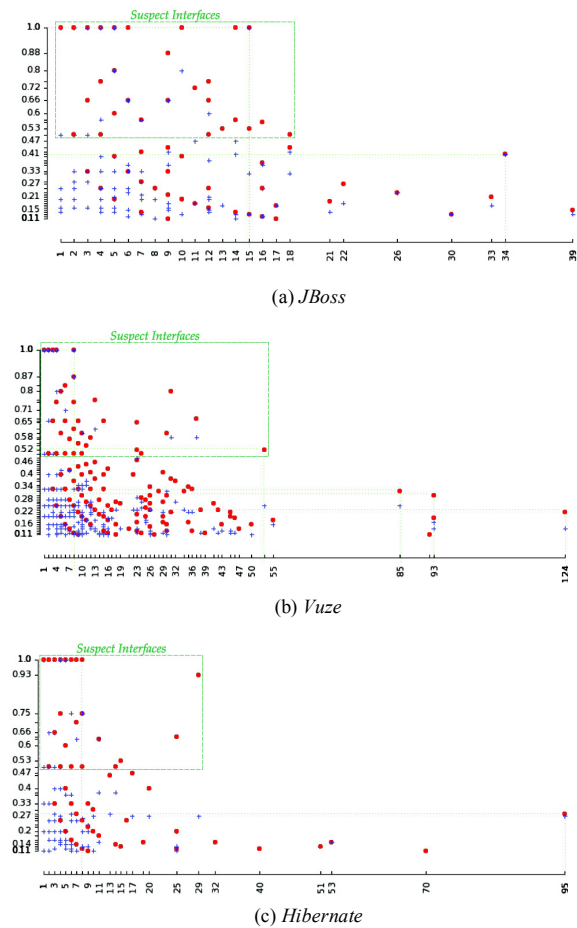

(a) *JBoss*

(b) *Vuze*

(c) *Hibernate*

Fig. 4 Scatter plot diagrams for IIS( $i$ ) and IIC( $i$ ) values for all interfaces in *JBoss, Vuze* and *Hibernate* applications.

4) *Some interfaces are completely duplicated:* looking to the most top of diagrams in Fig. 4 , where the IIS (and IIC) value is 1, we see that there are blue crosses (and/or red dots) at that line. This means that several interfaces in the studied applications are identical or those interfaces are completely duplicated in different interfaces. For example, Fig. 4 shows that there is an interface declaring 15 methods ( *i.e.,* its size is 15) and its completely duplicated in other interfaces (IIC = 1) and there is another interface that is identical to that one. By inspecting the *JBoss* code, we found that the interfaces *HAManagementServiceMBean* (in *management::j2ee::cluster* namespace) and *MEJB* (in *management::mejb* namespace) are those interfaces that declare 15 methods and are completely identical, in terms of public methods they declare. We also found two interfaces named *LocalJBossServerDomainMBean* in two different namespaces, *management::j2ee::deployers* and *management::j2ee*, where that in the later namespece declares 34 methods and completely clone *j2ee::deployers::LocalJBossServerDomainMBean* interface methods (14 methods in total). In fact, this hard coded similarity/clone between those two interfaces in *management::j2ee::deployers* and *management::j2ee* can be removed and replaced by a clean code structure if the *LocalJBossServerDomainMBean* in the later namespace extends that in the former one. In other terms, this unnecessarily similarity and method clones can be replaced by a good design specifying clearly *LocalJBossServerDomainMBean* in *j2ee* is a subtype of that in *j2ee::deployers*.

*C. IRIH Metric*

Fig. 5 shows the spread of IRIH values over the interfaces of *JBoss* 5(a), *Vuze* 5(b) and *Hibernate* 5(c). The values of IRIH are displayed with regard to the size of Interface sub-hierarchies. The figure shows that the redundancy of sub-typing relations is present in small hierarchies as in large ones.

Figures 5(a), 5(b) & 5(c) show that many interfaces, having small sub-hierarchy ($|subH(i)| < 15$), have significant values of IRIH. For example, in *Vuze* application Fig. 5(b), some interface with a sub-hierarchy of size 9, have a relatively large value of IRIH (0.55): *i.e.,* 55% of sub-typing relations in the interface sub-hierarchy are redundant.

In fact, that interface is precisely *common::IUserInterface*. This interface is implemented by four classes: (1) *common::UITemplate*, (2) *common::UITemplateHeadless*, (3) *common::UI* and (4) *telnet::UI*. All of *common::UITemplateHeadless*, *common::UI* and *telnet::UI* are sub-classes of *common::UITemplate*. Thus they are by default sub-types of *common::IUserInterface* interface. As a consequence, the static dependencies of those three classes to *common::IUserInterface* interface are redundant.

Many similar cases exist in *JBoss* and *Hibernate*, as illustrated in our examples in Section IV (Fig. 2). Figures 5(a), 5(b) & 5(c) show that for a large subset of interfaces with large sub-hierarchy ($|subH(i)| > 20$), the values of IRIH are relatively large. This is due to the fact that the hierarchies of those interfaces are composed of smaller hierarchies having significant redundancy of sub-typing relations. Hence, it is always preferable to start the refactoring process by interfaces having small sub-hierarchies.

A conclusion is that, the multiple inheritance mechanism via interfaces is really abused by software programmers. Software engineers should be more careful when stating needless static dependencies among classes and interfaces. They should take into account the complexity they add to interface hierarchies, and the static coupling they add among different software entities (classes, interfaces and packages).

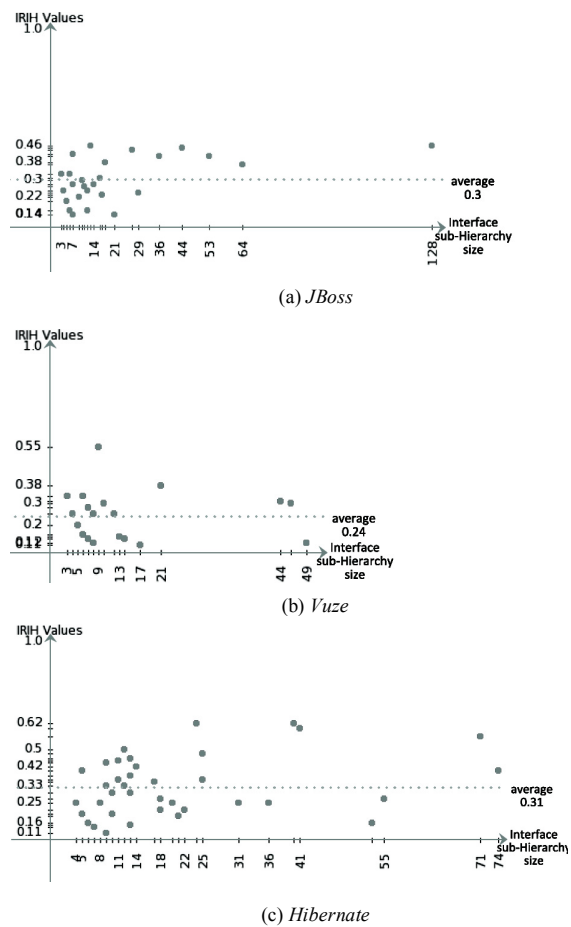

Fig. 5 Scatter plot diagrams for IRIH values in *JBoss*, *Vuze* and *Hibernate* applications.

## VI. RELATED WORKS

Recently, there has been a great progress in automatic detection of code bad smells and in automatic software refactoring [12], [20], [24], [17], [21]. Mens and Tourwe´ [19] surveyed existing approaches of software refactoring, show that existing software refactoring approaches are mainly based on code metrics and predefined bad smells in source code [6], [14], [3], [12].

However, on one hand, in spit of the well performance of existing metrics [6], they unfortunately do not address the particularities of interfaces [23]. On the other hand, Fowler and Beck [12] propose a set of bad smells in OO class design: *e.g.*, data class, god class, feature envy, duplicated code. They also propose refactorings for improving code quality with respect to the type of code smell. Following Fowler and Beck's definitions of class smells, Trifu and Marinescu establish a clear distinction between OO structural problems and code smells, and present a causal approach to restructuring OO applications [24]. Furthermore, Liu et al. [17] provided a deep analysis of the relationships among different kinds of bad smells and their influence on detection and resolution sequences. Unfortunately, none of those code smells and OO metrics attempt to address the particularities of interfaces.

In the literature, few recent works attempt to address the particularities of interfaces. Boxall and Araban define a set of primitive counting metrics to measure the complexity and usage of interfaces [4].

As for interface design quality, Martin proposed two design principles of interfaces [18]: the Interface Segregation Principle (ISP: "do not design fat interfaces"); and the Dependency Inversion Principle (DIP: "program to an interface, not an implementation"). According to the ISP, interfaces must be designed with regard to their clients: if an interface declares methods that are not used together, by the same set of clients, then that interface is a candidate to be split into smaller interfaces. Ideally, an interface should not expose any services that are not used by all its clients, and all the services of an interface should be used by every client of the interface. With regard to the ISP principle, Romano and Pinzger [23] used the Service Interface Usage Cohesion (SIUC) to measure the violation of ISP in Java interfaces. SIUC is defined by Perepletchikov [22], and it states that an interface has a strong cohesion if every client class of that interface uses all the methods declared in it. Romano and Pinzger conclude that in order to limit changes propagation and facilitate software maintenance, the ISP should be respected when designing interfaces.

Besides Martin's design principles ISP and DIP [18], Boxall and Araban's primitive metrics [4], and Perepletchikov's SIUC metric [22], there are no tools that help estimate the design quality and detect design anomalies in the design of Java interfaces. Our paper provides software developers with a complementary set of metrics that assist in detecting interface clones, quantifying similarities between interfaces and assessing interface subhierarchy. On one hand, IIS and IIC metrics help in assessing the organization of APIs and/or services into interfaces and the reusability among software interfaces. On another hand, IRIH outlines the unnecessary complexity within interface subhierarchy.

## VII. CONCLUSION AND FUTURE WORK

In this paper, we identified and explained via real examples two types of design defects specific to interfaces:

- Service clone at interfaces (Interface Similarity), inspired by the code clone bad smell.
- Redundancy in interface sub-hierarchies, inspired by the well-known principle so called DRY.

We also proposed metrics that measure the quality of interface design with respect to each design anomaly. We empirically investigated the interface design with our proposed design anomalies and metrics. The result of our study with three well-known Java applications (*JBoss, Vuze* and *Hibernate*) showed the following:

- The design anomalies we proposed are bad symptoms of interface design. They are present, to different degrees, at interfaces.
- Our metrics help in detecting bad designed inter-faces and extracting candidate interfaces for refactoring.

Our paper is a starting point for studying the quality of interface design and the impact of interface design defects on the quality of software application. Our findings implicate researchers and engineers should distinguish classes and interfaces when estimating the quality of software applications. Software engineers must do not create needless sub-typing relations to interfaces. Furthermore, they have to consider the reusability of interfaces and to be more careful while organizing service declarations into interfaces. They should use the mechanism of sub-typing between interfaces instead of hard copying method declarations between interfaces and designing similar interfaces.

As a future work, we plan to evaluate our metrics with more software systems. We plan to investigate the behaviour of our metrics with other OO metrics, such as CK metrics. We plan to identify the associations between interface design defects and those of classes.

### ACKNOWLEDGMENT

This publication was made possible by NPRP grant #09-1205-2-470 from the Qatar National Research Fund (a member of Qatar Foundation). The statements made herein are solely the responsibility of the author.

### REFERENCES

[1] H. Abdeen, S. Ducasse, H. A. Sahraoui, and I. Alloui. "Automatic package coupling and cycle minimization," in Proceedings of WCRE'09, pages 103–112. IEEE Computer Society Press, 2009.

[2] M. Balint, T. G¨ırba, and R. Marinescu. "How developers copy. In Proceedings of ICPC'06," pages 56–65, 2006.

[3] V. R. Basili, L. C. Briand, and W. L. Melo. "A validation of object-oriented design metrics as quality indicators," IEEE TSE, 22(10):751–761, 1996.

[4] M. A. S. Boxall and S. Araban. "Interface metrics for reusability analysis of components," in Proceedings of ASWEC'04, pages 40–51. IEEE Computer Society, 2004.

[5]  G. Bracha and W. Cook. "Mixin-based inheritance," in Proceedings of OOPSLA/ECOOP'90, ACM SIGPLAN Notices, volume 25, pages 303–311, Oct. 1990.

[6]  S. R. Chidamber and C. F. Kemerer. "A metrics suite for object oriented design," IEEE TSE, 20(6):476–493, June 1994.

[7]  D. Dig and R. Johnson. "The role of refactorings in api evolution," in Proceedings of ICSM '05, pages 389–398, 2005.

[8]  S. Ducasse, M. Lanza, and S. Tichelaar. "Moose: an Extensible Language-Independent Environment for Reengineering Object-Oriented Systems," in Proceedings of CoSET'00, June 2000.

[9]  S. Ducasse, O. Nierstrasz, N. Scha¨rli, R. Wuyts, and A. P. Black. "Traits: A mechanism for fine-grained reuse," ACM TPLAS, 28(2):331–388, Mar. 2006.

[10]  S. Eick, T. Graves, A. Karr, J. Marron, and A. Mockus. "Does code decay? assessing the evidence from change management data," IEEE TSE, 27(1):1–12, 2001.

[11]  J. R. Falleri, S. Denier, J. Laval, P. Vismara, and S. Ducasse. "Efficient retrieval and ranking of undesired package cycles in large software systems," in Proceedings of TOOLS'11, June 2011.

[12]  M. Fowler, K. Beck, J. Brant, W. Opdyke, and D. Roberts. "Refactoring: Improving the Design of Existing Code," Addison Wesley, 1999.

[13]  E. Gamma, R. Helm, R. Johnson, and J. Vlissides. "Design Patterns: Elements of Reusable Object-Oriented Software," Addison-Wesley Professional, 1995.

[14]  B. Henderson-Sellers. "Object-Oriented Metrics: Measures of Complexity," Prentice-Hall, 1996.

[15]  A. Hunt and D. Thomas. "The Pragmatic Programmer," Addison Wesley, 2000.

[16]  M. Lehman and L. Belady. "Program Evolution: Processes of Software Change," London Academic Press, London, 1985.

[17]  H. Liu, Z. Ma, W. Shao, and Z. Niu. "Schedule of bad smell detection and resolution: A new way to save effort," IEEE TSE, 38(1):220–235, Jan. 2012.

[18]  R. C. Martin. "Design principles and design patterns," 2000.

[19]  T. Mens and T. Tourwe´. "A survey of software refactoring," IEEE TSE, 30(2):126–138, 2004.

[20]  M. J. Munro. "Product metrics for automatic identification of bad smell design problems in java source-code," in Proceedings of METRICS'05, pages 15–23. IEEE Computer Society, 2005.

[21]  E. Murphy-Hill, C. Parnin, and A. P. Black. "How we refactor, and how we know it," IEEE TSE, 38:5–18, 2012.

[22]  M. Perepletchikov, C. Ryan, and K. Frampton. "Cohesion metrics for predicting maintainability of service-oriented software," in Proceedings of QSIC '07, pages 328–335. 2007.

[23]  D. Romano and M. Pinzger. "Using source code metrics to predict change-prone java interfaces," in Proceedings of ICSM'11, pages 303–312, 2011.

[24]  A. Trifu and R. Marinescu. "Diagnosing design problems in object oriented systems," in Proceedings of WCRE'05, pages 155–164. IEEE Computer Society, 2005.

[25]  B. Venners. "Designing with interfaces," 1998.

[26]  N. Warren and P. Bishop. "Java in Practice. Addison Wesley," 1999.